\documentclass[a4paper, 11pt]{article}

\usepackage{graphicx}
\usepackage{epsf,amsmath,bbold,amsfonts,stmaryrd}

\usepackage[utf8]{inputenc}
\usepackage{mathrsfs}
\usepackage{appendix}
\usepackage{amssymb}
\usepackage{float}
\usepackage{color}
\usepackage{cite}
\usepackage{hyperref}
\hypersetup{pageanchor=false}
\usepackage{indentfirst}
\usepackage{url}
\usepackage{float}
\usepackage{caption}
\usepackage[numbers,square,comma,sort&compress,merge]{natbib}

\usepackage{subfigure}

\usepackage{ulem}

\def\nn{\nonumber}

\def\te{\tilde{\eta}}
\def\tx{\tilde{\xi}}

\def\be{\begin{equation}}
\def\ee{\end{equation}}

\def\bea{\begin{eqnarray}}
\def\eea{\end{eqnarray}}

\def\ba{\begin{array}}
\def\ea{\end{array}}

\def\bc{\begin{center}}
\def\ec{\end{center}}

\def\bl{\begin{flushleft}}
\def\el{\end{flushleft}}

\def\br{\begin{flushright}}
\def\er{\end{flushright}}

\def\bi{\begin{itemize}}
\def\ei{\end{itemize}}

\def\bt{\begin{tabular}}
\def\et{\end{tabular}}

\newcommand{\xmi}{x_\text{min}}
\newcommand{\xma}{x_\text{max}}
\newcommand{\ymi}{y_\text{min}}
\newcommand{\yma}{y_\text{max}}

\numberwithin{equation}{section}

\begin{document}
\title{How different are shadows of compact objects with and without horizons?}

\author{
Xiangyu Wang$^{1}$, Yehui Hou$^{2}$, Minyong Guo$^{1\ast}$}
\date{}

\maketitle

\vspace{-10mm}

\begin{center}
{\it
$^1$ Department of Physics, Beijing Normal University,
Beijing 100875, P. R. China\\\vspace{4mm}

$^2$Department of Physics, Peking University, No.5 Yiheyuan Rd, Beijing
100871, P.R. China\\\vspace{4mm}
}
\end{center}

\vspace{8mm}

\begin{abstract}
In this work, we theoretically assume that a compact object (CO) has a dark surface such that this simplified CO has no emissions and no reflections. 
Considering that the radius of the surface can be located inside or outside the photon region, which is closely related to the shadow curve, we investigate whether a CO without an event horizon can produce shadow structures similar to those of black holes and compare the shadows of COs with and without horizons. 
In particular, by introducing the (possible) observational photon region, we analytically construct an exact correspondence between the shadow curves and the impact parameters of photons; 
we find that there are indeed several differences between the shadows of COs without horizons and those of black holes. 
More precisely, we find that the shadow curve is still determined by the photon region when the radius of the surface is small enough to retain a whole photon region outside the shell. When only part of the photon region remains, the shadow curve is partially determined by the photon region, and the remaining portion of the shadow curve is partly controlled by the impact parameters of photons that have a turning point on the surface. When there is no photon region outside the surface, the shadow curve is totally controlled by the impact parameters of photons, which have a turning point on the surface.

\end{abstract}

\vfill{\footnotesize $\ast$ Corresponding author: minyongguo@bnu.edu.cn}

\maketitle

\newpage
\baselineskip 18pt
\section{Introduction}\label{sec1}
It is known that due to the strong gravitational field around a black hole, light has to bend and form a central dark area in the view of distant observers, dubbed the black hole shadow. For black hole shadows, one of the most apparent features might be the so-called shadow curve (also referred to as the critical curve in the literature \cite{Gralla:2019xty, Peng:2020wun}). In most cases, we know that the shadow curve is closely related to the photon region, which is composed of the spherical photon orbits \footnote{The spherical photon orbits are usually defined by $r=\text{const}$ in a stationary and axisymmetric spacetime, where $r$ is the radial coordinate. In a curved spacetime as a radial parameter, $r=\text{const}$ generally does not imply the spherical meaning in flat space. A stricter definition can be found in \cite{Cunha:2017eoe}, where the authors introduced a new terminology: the fundamental photon orbits. Some related works concerned with fundamental photon orbits can be seen in \cite{Li:2021zct, Guo:2020qwk}.}, even though the essence of a black hole shadow is the existence of an event horizon that can capture photons with specific impact parameters.

In recent years, the central depression of emissions has been found in black hole images photographed by the Event Horizon Telescope (EHT) \cite{EventHorizonTelescope:2022xqj, EventHorizonTelescope:2022ndl, EventHorizonTelescope:2019ggy, EventHorizonTelescope:2022ppi, EventHorizonTelescope:2022ago, EventHorizonTelescope:2022okn, EventHorizonTelescope:2022gsd}. There have been many exciting works on shadows in terms of the EHT \cite{Hou:2021okc, Guo:2020zmf, Zhu:2023btl, Chen:2022scf, Atamurotov:2022iwj, Sau:2022afl, Vagnozzi:2019apd, Grenzebach:2014fha, Wei:2013kza, Perlick:2018iye, Zeng:2020dco, Li:2020drn, Wang:2018eui, Guo:2018kis, Moffat:2019uxp, Huang:2016qnl, Hu:2022sej, Hou:2022gge, Wen:2022hkv, Sengo:2022jif, Chen:2022qrw, He:2021htq, Zhang:2020xub, Junior:2021svb, Tang:2022bcm, Meng:2022kjs, Li:2021ypw, Guo:2020qxy, Wu:2022ydc, Vagnozzi:2022moj, Gan:2021xdl, Wang:2020emr, Olivares:2018abq, Kleihaus:2005me, Kleihaus:2007vk, Herdeiro:2015gia, Siemonsen:2020hcg, Herdeiro:2021lwl, Vincent:2015xta, Allahyari:2019jqz, Khodadi:2020jij, Roy:2021uye, Chen:2022nbb, Afrin:2022ztr}, among which were investigations into whether some specific compact objects (COs) without horizons could mimic black hole shadows \cite{Olivares:2018abq, Kleihaus:2005me, Kleihaus:2007vk, Herdeiro:2015gia, Siemonsen:2020hcg, Herdeiro:2021lwl, Vincent:2015xta, Kumar:2020ltt, KumarWalia:2022aop}, that is, if the shadow is a sufficient condition for the existence of an event horizon. Along this line, previous studies have mainly focused on boson stars, which have no hard emitting surface. Considering that boson stars are illuminated by the surrounding accretion flows that have a cutoff in luminance
at the inner edge of the accretion disk, the authors have numerically found that some boson stars, especially Proca stars, can produce images including shadow structures similar to black holes.

In our work, we consider a CO with a surface and theoretically investigate the difference between the shadows of COs with and without horizons. For simplicity, we focus on a model with two ideal assumptions. Compared with the assumptions of luminous accretion flows or other light sources in the background, we first assume that CO is a nonluminous body; that is, the surface of CO has no emissions. 
Second, we assert that the CO is a dark star so that little light reflects from the surface of the CO. 
Thus, the reflections can be omitted. In short, in our simplified model, the CO does not transmit and does not reflect lights, thus behaving effectively like an event horizon. However, compared to that of a black hole, the radius of the surface of the CO can be chosen arbitrarily, while the event horizon is fixed.
Moreover, since the radius of the surface is not fixed, there might be no photon region, or only part of the photon region remains outside the surface of the CO. As we know, the black hole shadow curve is usually determined by the photon region. Thus, it is fascinating to theoretically study the shadow structures of the CO in our model. It has been shown that there are several types of COs in general relativity, including constant-density stars \cite{1983Black}, thin-shell gravastars \cite{Pani:2009ss}, boson stars \cite{Olivares:2018abq, Kleihaus:2005me}, Proca stars \cite{Brito:2015pxa} and so on. In this work, for this purpose, we focus on a rotating and horizonless body to preserve a photon shell. On the other hand, for convenience, we want to investigate within an analytic metric. However, such an exact metric has not been found up to now. Note that the Lense–Thirring metric is a slow-rotation large-distance approximation to the gravitational field outside a massive rotating body, that is to say, the Lense-Thirring metric is an excellent approximation to the exterior spacetime geometry $r>r_s$, when $r^2_s\gg J$, where $r_s$ is the surface radius and $J$ is the momentum of the slow rotating body \cite{mashhoon1984influence}. Thus, in this work, we pay attention to the Painlev\'e-Gullstrand form of the Lense-Thirring spacetime proposed recently in \cite{Baines:2020unr} and focus on the region at $r>r_s$ by imposing $r^2_s\gg J$.

The remaining parts of this paper are organized as follows: In sec. \ref{sec2}, we review the Painlev\'e-Gullstrand form of the Lense-Thirring spacetime, and we discuss the geodesics in sec. \ref{sec3}. We introduce the (possible) observational photon region and have a detailed study of the shadow curves for COs with and without horizons. The main conclusions are summarized in sec. \ref{sec4}. In this work, we set the fundamental constants $c$ and $G$, and we work in the signature convention $(-, +, +, +)$ for the spacetime metric.

\section{Painlev\'e-Gullstrand form of the Lense-Thirring spacetime}\label{sec2}
Since we use the Lense-Thirring metric to model a horizonless CO, we review the Lense-Thirring spacetime.
\subsection{Metric}
In 1918, Lense and Tirring proposed an approximate solution to describe a slow rotating large-distance stationary isolated body in the framework of the vacuum Einstein equations \cite{mashhoon1984influence}, which takes
\bea\label{orfo}
ds^2=&-&\left[1-\frac{2M}{r}+\mathcal{O}\left(\frac{1}{r^2}\right)\right]dt^2-\left[\frac{4J\sin^2\theta}{r}+\mathcal{O}\left(\frac{1}{r^2}\right)\right]d\phi dt\nn\\
&+&\left[1+\frac{2M}{r}+\mathcal{O}\left(\frac{1}{r^2}\right)\right]\left[dr^2+r^2\left(d\theta^2+\sin^2\theta d\phi^2\right)\right]\,
\eea
where $M$ and $J$ are the mass and the angular momentum, respectively. $\mathcal{O}(r^{-2}) $ denotes the subdominant terms. By properly regulating the specific forms of $\mathcal{O}(r^{-2})$, one can obtain various metrics with the same asymptotic limit at large distances, which are physically different from each other. Recently, Baines et al. constructed an explicit Painlev\'e-Gullstrand variant of the --Lense-Thirring spacetime \cite{Baines:2020unr}, for which the metric is
\bea\label{metric}
ds^2=-dt^2+\left(dr+\sqrt{\frac{2M}{r}}dt\right)^2+r^2\left[d\theta^2+\sin^2\theta\left(d\phi-\frac{2J}{r^3}dt\right)^2\right]\,.
\eea
There are three solid advantages for this new version of the --Lense-Thirring spacetime, of which the first is that the metric reduces to the Painlev\'e--Gullstrand version of the Schwarzschild black hole solution when $J=0$; the second is that the azimuthal dependence takes a partial Painlev\'e-Gullstrand form, that is, $g_{\phi\phi}(d\phi-v^\phi dt)^2=g_{\phi\phi}(d\phi-\omega dt)^2$, where $v^\phi$ is minus the azimuthal component of the shift vector in the ADM formalism denoting the ``flow '' of the space in the azimuthal direction and $\omega=g_{t\phi}/g_{\phi\phi}$ is the angular velocity of the spacetime; and the third is that all the spatial dependence is in exact Painlev\'e-Gullstrand type form, which implies that the spatial hypersurface $t=\text{const}$ is flat. These exciting features make the Painlev\'e-Gullstrand variant much easier to calculate with respect to the tetrads, the curvature components, and the geodesic analysis than any other variant of the Lense-Thirring spacetime \cite{Baines:2021qaw, Baines:2021qfm}.

On the other hand, from the original asymptotic form in Eq. (\ref{orfo}), we can see that the Lense-Thirring metric should make sense only in the region $r>r_s$. The metric in Eq. (\ref{orfo}) has a coordinate singularity $r=2M$ when neglecting the subdominant terms so that the Lense-Thirring spacetime should be valid when the condition $r_s>2M$ holds. Moreover, for a slowly rotating object, we must have $J/r_s^2\ll1$. Thus, we should also impose the conditions $J/r_s^2\ll1, r_s>2M$ on the Painlev\'e-Gullstrand version of the Lense-Thirring spacetime when investigating the properties of the Painlev\'e-Gullstrand form.

\subsection{Geodesics}
In this subsection, we review the geodesics in the Painlev\'e-Gullstrand form of the Lense-Thirring spacetime, which has been carefully studied in \cite{Baines:2021qfm}. Similar to the Kerr spacetime, there are also four conserved quantities along the geodesics of free particles: the mass $m$, the energy $E$, the axial angular momentum $L$, and the Carter constant $C$. For simplicity and without loss of generality, we set $m=0$ for photons and $m=1$ for time-like particles. Then, the four-momentum $p^a$ reads
\bea
p^a=\dot{t}\left(\frac{\partial}{\partial t}\right)^a+\dot{r}\left(\frac{\partial}{\partial r}\right)^a+\dot{\theta}\left(\frac{\partial}{\partial \theta}\right)^a+\dot{\phi}\left(\frac{\partial}{\partial \phi}\right)^a\,,
\eea
with `` $\dot{}$ '' denoting the derivative with respect to the affine parameter $\tau$. Considering $p^a p_a=0$ for photons and $p^a p_a=-1$ for time-like particles, $\tau$ can be seen as the proper time for time-like worldliness. Then, the conserved quantities $E, L, C$ can be written as
\bea
&E&=-p_t=\left(1-\frac{2M}{r}-\frac{4J^2\sin^2\theta}{r^4}\right)\dot{t}-\sqrt{\frac{2M}{r}}\dot{r}+\frac{2J\sin^2\theta}{r}\dot{\phi}\,,\nn\\
&L&=p_\phi=r^2\sin^2\theta\left(\dot{\phi}-\frac{2J}{r^3}\dot{t}\right)\,,\quad C=r^4\dot{\theta}^2+\frac{L^2}{\sin^2\theta}\,,
\eea
explicitly. For time-like particles, $E$ and $L$ can now be treated as the energy per unit mass and the angular momentum per unit mass. Then, combined with the condition $-p^a p_a=m\in\{0, 1\}$, one can obtain the exact expressions of the components of the four momentum $p^a$ as follows:
\bea\label{geeq}
\dot{r}&=&S_r\sqrt{R(r)}\,,\nn\\
\dot{t}&=&\frac{E-2JL/r^3+S_r\sqrt{(2M/r)R(r)}}{(1-2M/r)}\,,\nn\\
\dot{\theta}&=&S_\theta\frac{\sqrt{\Theta(\theta)}}{r^2}\,,\nn\\
\dot{\phi}&=&\frac{L}{r^2\sin^2\theta}+2J\frac{E-2JL/r^3+S_\phi\sqrt{(2M/r)R(r)}}{r^3(1-2M/r)}\,,
\eea
where we define
\bea
R(r)&=&\left(E-\frac{2JL}{r^3}\right)^2-\left(m+\frac{C}{r^2}\right)\left(1-\frac{2M}{r}\right)\,,\\
\Theta(\theta)&=&C-\frac{L^2}{\sin^2\theta}\,,
\eea
as the effective potential functions governing the radial and polar motions, and
\bea
S_r&=&
\begin{cases}\label{sign}
+1\:\:\:\text{outgoing geodesic}\\
- 1\:\:\:\text{ingoing geodesic}
\end{cases}\;\nn\\
S_{\theta}&=&
\begin{cases}
+1\:\:\:\text{increasing declination geodesic}\\
- 1\:\:\:\text{decreasing declination geodesic}
\end{cases}\;\nn\\
S_{\phi
}&=&
\begin{cases}
+1\:\:\:\text{prograde geodesic}\\
- 1\:\:\:\text{retrograde geodesic}
\end{cases}\,
\eea
following the conventions in \cite{Baines:2021qfm}. The context for each equation in Eq. (\ref{sign}) denotes the corresponding physical interpretation. Here, we want to stress that $S_r$ and $S_\phi$ appear separately in the $t$-motion and $\phi$-motion due to the Painlev\'e-Gullstrand form; however, for geodesic equations of Kerr spacetime in Boyer-Lindquist coordinates, $S_r$ comes up only in the radial motion, and $S_\phi$ is not necessarily introduced. Then, one can explore the properties of null and time-like geodesics by adequately manipulating the equations in (\ref{geeq}).

\section{Observational photon region and shadow curve}\label{sec3}
This section focuses on the photon region and shadow curve in the Painlev\'e-Gullstrand form of the Lense-Thirring spacetime. Considering that the null orbits are independent of photon energies, it is convenient to introduce the impact parameters
\bea
\xi=\frac{L}{E}\,, \quad\quad\eta=\frac{C-L^2}{E^2}\,.
\eea
to characterize the photon orbits. The conditions can determine the photon region
\bea
R(r)=\partial_rR(r)=0\,,
\eea
which gives us the expressions of the impact parameters in terms of the radius,
\bea
\tilde{\xi}&=&\frac{-3M\tilde{r}^3+\tilde{r}^4}{2J(3M-2\tilde{r})}\,, \nn\\
\tilde{\eta}&=&-\frac{\tilde{r}^3[\tilde{r}^3(\tilde{r}-3M)^2+36J^2(\tilde{r}-2M)]}{4J^2(3M-2\tilde{r})^2}\,.
\eea
We use $\tilde{r}$ to denote the radius of the photon orbit in the photon region, and $\tilde{\xi}, \tilde{\eta}$ are the corresponding impact parameters. Furthermore, from $\tilde{\eta}=0$, we can obtain two roots $r_{p-}<r_{p+}$ in the region $\tilde{r}>2M$, which implies that the radial range of the photon region is
\be
\tilde{r}\in[r_{p-},r_{p+}]\,.
\ee
We note that $r_{p\pm}$ cannot be analytically given in general; however, when $J\to0$, one can find \cite{Baines:2021qfm}
\bea
r_{p\pm}=3M\pm\frac{2J}{\sqrt{3}M}+\mathcal{O}(J^2)\,.
\eea
Considering $r_s>2M$ for COs, in light of $r_{p\pm}$, we divide the range of $r_s$ into three parts, that is, (1) $2M<r_s<r_{p-}$, (2) $r_s>r_{p+}$, and (3) $r_{p-}<r_s<r_{p+}$, and we study the shadow curve for each case.

\subsection{Review of black hole shadows}\label{3s1}

\begin{figure}[h!]

\centering
\includegraphics[width=5.5in]{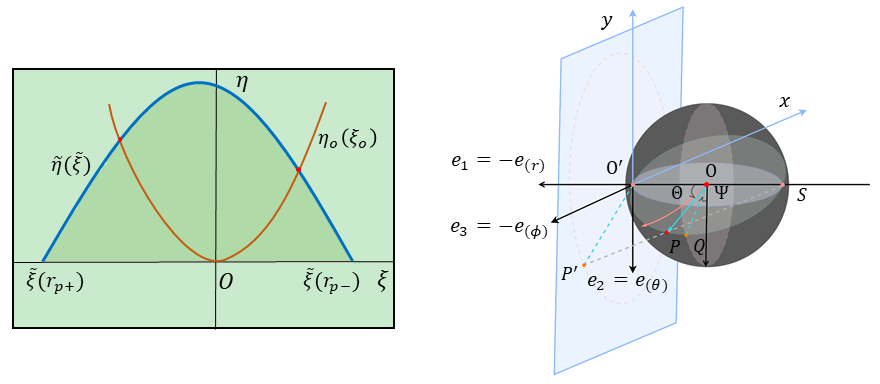}

\caption{An illustration of the observational photon region for a black hole in the $\xi O\eta$ plane is shown in the left panel. The right panel is borrowed from Fig. 11 of our previous work \cite{Hu:2020usx}, which presents the celestial coordinates $(\Theta, \Psi)$ and standard Cartesian
coordinates $(x, y)$ in the local rest frame of the observers. }\label{bhp}
\end{figure}

Before we discuss the shadows of the COs, we first review the shadows of ordinary black holes. To determine the shadow of a black hole, in addition to the photon region, there is a second condition related to the observational angle. For a certain observational angle $\theta_o$, we can see that the term under the square root $\Theta(\theta_o)\ge0$ must be satisfied in polar motion, which produces
\bea
\Theta(\theta_o)=\eta_o-\frac{\xi_o^2}{\sin^2\theta_o}\ge0\,,
\eea
and a new function $\eta_o(\xi_o)=\frac{\xi_o^2}{\sin^2\theta_o}$. That is, the photons can reach the observer if their impact parameters satisfy the above condition. 
Combining the critical impact parameters $\tilde{\eta}(\tilde{\xi})$ with the constraint $\Theta(\theta_o)\ge0$, one can fix the photons exactly that have critical impact parameters and those that can escape to observers if they are perturbed. 
As a result, the shadow curve is formed by these photons since the surface of the black hole, that is, the horizon, is always inside the photon region.

In the study of shadows of COs, including black holes, we find it convenient to define the observational photon region (OPR) and possible observational photon region (POPR). 
The OPR is defined as the set of impact parameters for which the photons with these impact parameters precisely determine the shadow curve for observers with a specific observational angle. 
The POPR is defined as the union of the OPRs at all possible observational angles. Thus, for the case of black holes, the POPR is composed of the critical impact parameters $\tilde{\eta}(\tilde{\xi})$, and the elements of the OPR are the critical impact parameters $\tilde{\eta}(\tilde{\xi})$, which also satisfy the condition $\Theta(\theta_o)\ge0$. In the left panel of Fig. \ref{bhp}, we present the functions of $\tilde{\eta}(\tilde{\xi})$ and $\eta_o(\xi_o)$ in the $\xi O\eta$ plane and find that the two functions have two intersections. The OPR corresponds to the segment of $\tilde{\eta}(\tilde{\xi})$ between the two intersections, and the POPR corresponds to a piece of $\tilde{\eta}(\tilde{\xi})$ above the $\xi$-axis.

Then, one can calculate the shadow curve by standard methods, that is, introducing the celestial coordinates and obtaining the projections on the screen of observers. In this work, we employ the stereographic projection method, which has been used in our previous work \cite{Hu:2020usx}. We also bring Fig. 11 from the work \cite{Hu:2020usx} into the right panel of Fig. \ref{bhp} to give a deep intuition on the celestial coordinates and Cartesian
coordinates $(x, y)$ in the local rest frame of the observers.

In terms of the metric in Eq. (\ref{metric}), the local rest frame of observers can be defined as
\bea
e_0&=&\hat{e}_{(t)}=\partial_t-\sqrt{\frac{2M}{r}}\partial_r+\frac{2J}{r^3}\partial_\phi\,,\\
e_1&=&-\hat{e}_{(r)}=-\partial_r\,,\\
e_2&=&\hat{e}_{(\theta)}=\frac{1}{r}\partial_\theta\,,\\
e_3&=&-\hat{e}_{(\phi)}=-\frac{1}{r\sin\theta}\partial_\phi\,.
\eea
It is not hard to verify that these bases are normalized and orthogonal to each other. Moreover, since $\hat{e}_{(t)}\cdot\partial_\phi=0$, the observer with the 4-velocity $\hat{u}=e_0$ in this local rest frame has zero angular momentum for infinity. Therefore, this frame is usually called the ZAMO reference frame. In our model, the relation between the celestial coordinates $(\Theta, \Psi)$ and the 4-momentum of the OPR takes
\bea\label{tp}
\Theta=\arccos\left(\sqrt{\frac{2M}{r_0}}+\frac{\dot{\tilde{r}}_o}{\dot{\tilde{t}}_o}\right)\,,\quad \Psi=-\arctan\left(\frac{\tilde{\xi}}{\sqrt{\tilde{\eta}\csc^2\theta_o-\tilde{\xi}^2}}\right)\,,
\eea
where `` $\sim$ '' means evaluated with critical impact parameters $\tilde{\xi}$ and $\tilde{\eta}$, and the subscript `` $_o$ '' means evaluated at the observer with coordinates $(0, r_o, \theta_o, 0)$. Then, the Cartesian coordinates $(x, y)$ on the screen can be defined as
\bea\label{xy}
x=-2\tan\frac{\Theta}{2}\sin\Psi\,,\quad y=-2\tan\frac{\Theta}{2}\cos\Psi\,,
\eea
where we have chosen the energy of the photon observed by the ZAMOs to be unity, considering that the trajectories of photons are independent of the energies.

\subsection{Shadows of COs without horizons}

In this subsection, we study the shadows of COs, which have no horizon. For simplicity, we assume that the COs are nonluminous bodies, and they neither transmit nor reflect light. We recall that the spacetime outside a CO that we consider in this work is modeled by the Painlev\'e-Gullstrand form of the Lense-Thirring spacetime, and we investigate the shadows in three situations, (1) $2M<r_s<r_{p-}$, (2) $r_s>r_{p+}$, and (3) $r_{p-}<r_s<r_{p+}$.

\begin{figure}[h!]

\centering
\includegraphics[width=6.5in]{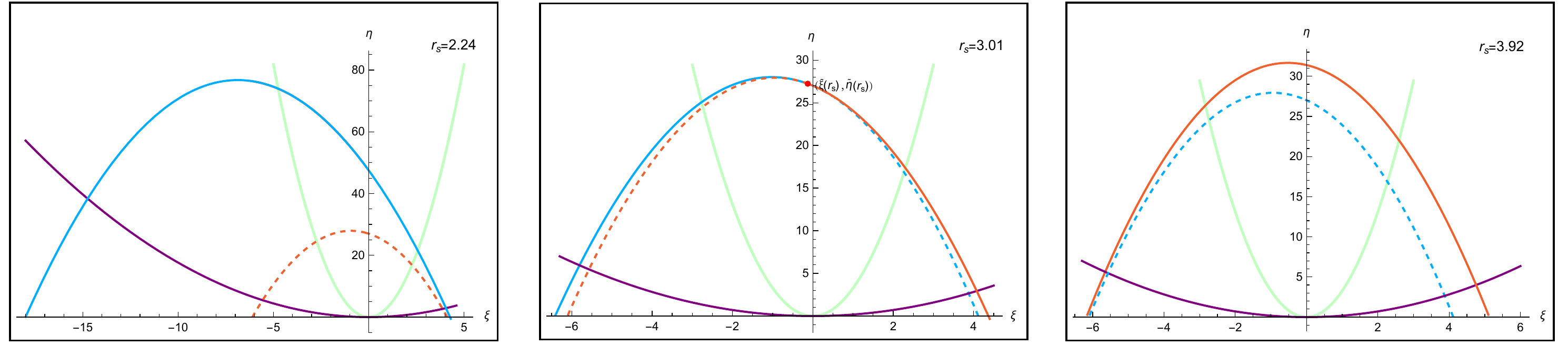}

\caption{Plots of the functions $\te(\tx)$, $\eta_s(\xi_s)$ and $\eta_o(\xi_o)$ in the $\xi O\eta$ plane for $r_s=2.24$, $r_s=3.01$ and $r_s=3.92$ with $M=1$ and $J=0.5$. In each plot, $\te(\tx)$ is shown in the dashed line, $\eta_s(\xi_s)$ is shown in the solid line with downward opening, $\eta_o(\xi_o)$ with $\theta_o=17^\circ$ is given by the green line and $\eta_o(\xi_o)$ with $\theta_o=80^\circ$ is given by the purple line. In addition, the POPR is shown by the red line in each plot, while the blue line has no contribution to the shadow curve.}\label{el}
\end{figure}

As mentioned above, the shadow is clear if we find the corresponding OPR. Thus, the main task is to look for the OPR for each case. Since the CO is regarded as a dark body in our work, the effect on lights is equivalent to that of the event horizon of a black hole; that is, the photons cannot return if they meet the surface of the CO. As a result, the incoming photons, which have two turning points in the radial motion, cannot escape to infinity if the outer turning point is inside the surface of the CO. Thus, if $r_s$ is not less than $\tilde{r}_{p-}$, the part of the photon region inside the surface of the CO has no contribution to the POPR. More precisely, from $R(r_s)=0$, we can obtain a new relation between $\xi_{s}$ and $\eta_{s}$ as follows:
\bea
\eta_s=-\frac{(r_s-2J\xi_s)^2}{(2M-r_s)r_s^3}-\xi_s^2\,,
\eea
where the subscript `` $s$ '' means evaluated at $r=r_s$. Considering that the radius of the surface $r_s$ can be the inner or outer turning point, which corresponds to different values of $(\xi_s, \eta_s)$, $\eta_s(\xi_s)$ becomes the new critical parameter when $r_s>\tilde{r}$, where $\tilde{r}$ is the radius of the photon region with $\tilde{\eta}(\tilde{\xi})$. In Fig. \ref{el}, we give examples of $\tilde{\eta}(\tilde{\xi})$, $\eta_s(\xi_s)$ and $\eta_o(\xi_o)$ for three cases at the observational angles $\theta_o=17^\circ$ and $\theta=80^\circ$ with the mass and the angular momentum of the CO chosen as $M=1$ and $J=0.5$ here and after this. By numerically solving the equation $\tilde{\eta}=0$, we find
\bea
r_{p-}\simeq2.47\,,\quad r_{p+}\simeq3.56\,.
\eea

\begin{figure}[h!]

\centering
\includegraphics[width=6.5in]{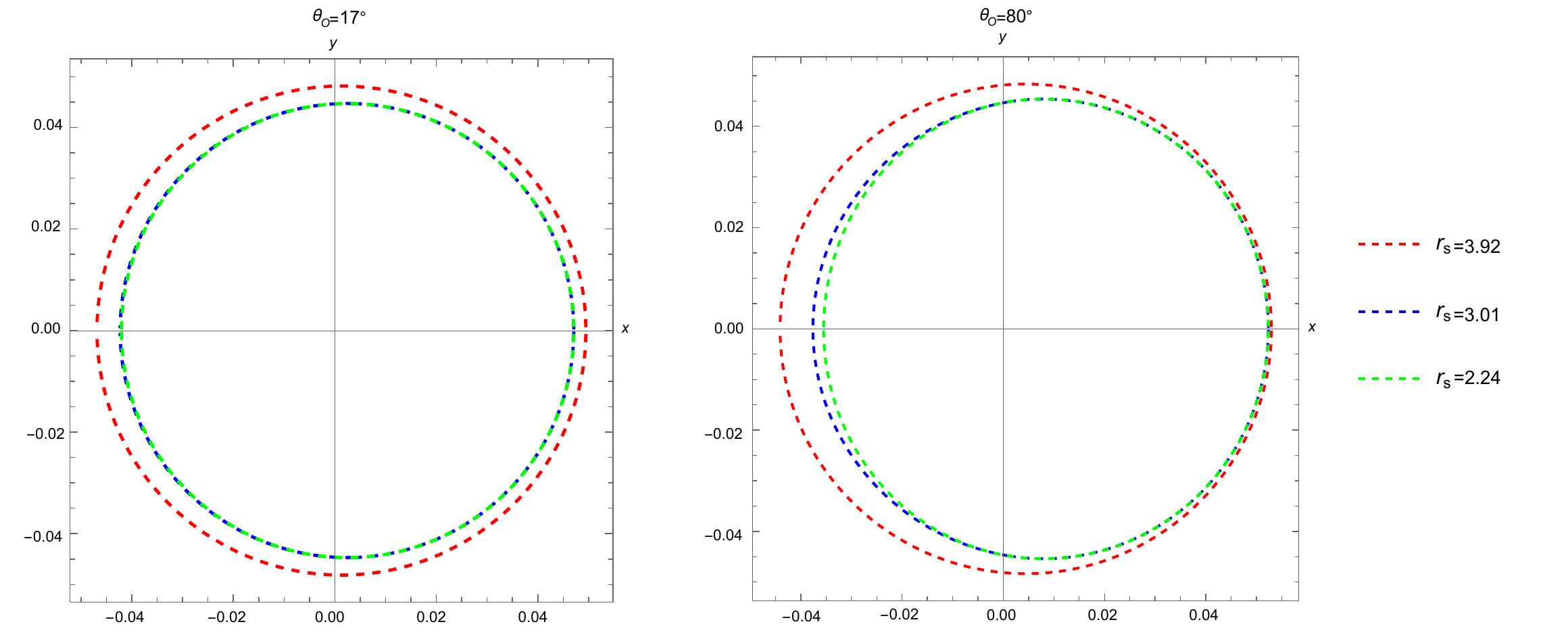}

\caption{Plots of shadow curves of COs. In the left plot, we set $\theta_o=17^\circ$, and in the right plot, we set $\theta_o=80^\circ$. In both plots, the green, blue and red lines denote the shadow curves with $r_s=2.24$, $r_s=3.01$, and $r_s=3.92$, respectively.}\label{asc}
\end{figure}

In addition, assuming $R=\partial_r R=\partial_r^2 R=0$ for prograde time-like particles
, we can find the radius of the innermost stable circular orbit $r_I\simeq4.29$. Considering that the horizon is at $r_h=2$, we set $r_s=\frac{r_h+r_{p-}}{2}\simeq2.24<r_{p-}$, $r_{p-}<r_s=\frac{r_{p-}+r_{p+}}{2}\simeq3.01<r_{p+}$ and $r_s=\frac{r_{p+}+r_I}{2}\simeq3.92>r_{p+}$ for the plots from left to right in Fig. \ref{el}. In addition, for each plot, the dashed line denotes $\tilde{\eta}(\tilde{\xi})$, the other curve with a downward opening indicated by a solid line denotes $\eta_s(\xi_s)$, the curve with an upward opening drawn in green is $\eta_o(\xi_o)$ with $\theta_o=17^\circ$, and the other curve with an upward opening drawn in purple is $\eta_o(\xi_o)$ with $\theta_o=80^\circ$. For the middle plot in Fig. \ref{el} with $r_{p-}<r_s<r_{p+}$, there is an intersection point $(\tx(r_s), \te(r_s))$ of $\tilde{\eta}(\tilde{\xi})$ and $\eta_s(\xi_s)$, which means that the two turning points of photons coincide with the radius $r=r_s$. When $\xi>\tx(r_s)$, we find that $r_s$ is the outer turning point of $R(r_s)=0$ and $r_s>\tilde{r}$. In contrast, when $\xi<\tx(r_s)$, we find that $r_s$ is the inner turning point of $R(r_s)=0$ and $r_s<\tilde{r}$. Therefore, the red line is the POPR. The impact parameters that are not in POPR are shown in blue. Moreover, combined with the condition from the observer at $\theta_o=17^\circ$ ($\theta_o=80^\circ$), the POR is the segment of the red line between the intersections of the red and green (purple) lines. For the left plot in Fig. \ref{el} with $r_s<r_{p-}$, we can see that the POPR is still determined by $\te(\tx)$, which is the same as that in black hole spacetime since the surface of the CO is always hidden in the photon region. The OPR is the segment of $\te(\tx)$ between the intersections of the red line $\te(\tx)$ and the green line $\eta_o(\xi_o)$. For the right plot in Fig. \ref{el} with $r_s>r_{p+}$, we can see that the POPR is determined by the solid line $\eta_s(\xi_s)$ since the photon region is completely encapsulated by the surface of the CO. The OPR is now given by the segment of the red line $\eta_s(\xi_s)$ between the intersections of $\eta_s(\xi_s)$ and $\eta_o(\xi_o)$.

\begin{figure}[h!]

\centering
\includegraphics[width=3in]{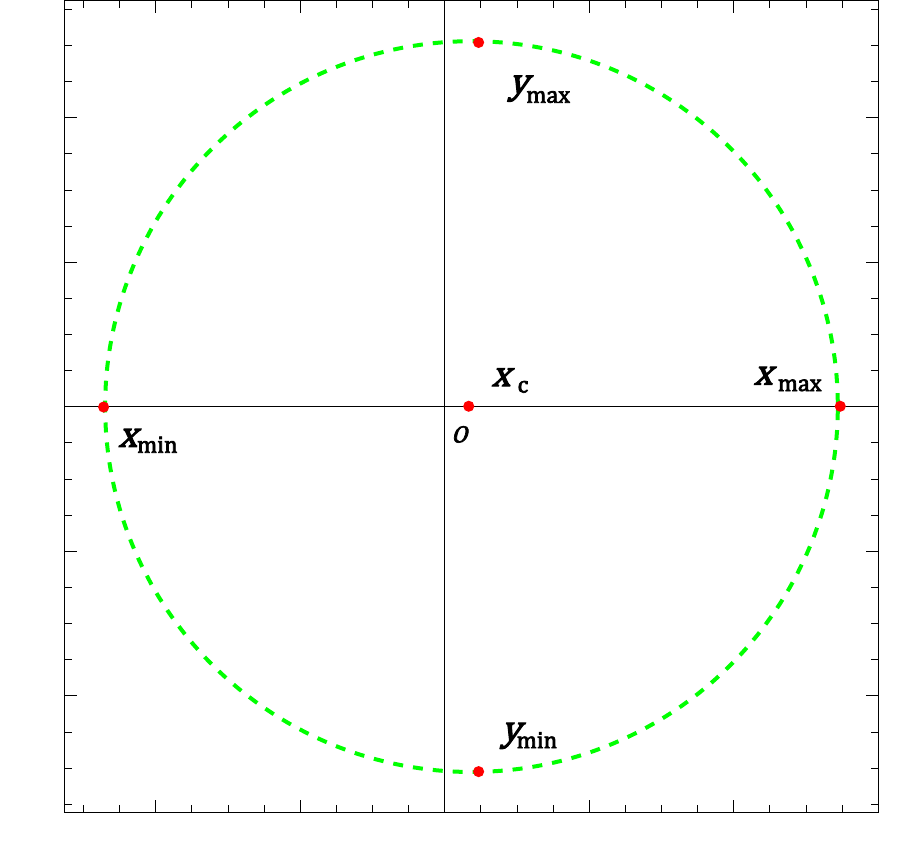}

\caption{An illustration of the coordinates of the points at which the shadow curve intersects the two axes on the screen.}\label{dia}
\end{figure}

Then, the shadows of COs without horizons can be calculated with the help of Eqs. (\ref{tp}) and (\ref{xy}). In Fig. \ref{asc}, we show the shadow curves with dashed lines at $\theta_o=17^\circ$ for the left plot and $\theta=80^\circ$ for the right plot. The red, blue and green lines correspond to $r_s=3.92>r_{p+}$, $r_{p-}<r_s=3.01<r_{p+}$ and $r_s=2.24<r_{p-}$, respectively. As we have discussed above, the shadow curve is exactly determined by the OPR, and we note that in Fig. \ref{el}, the dashed line in each plot represents the same photon region, that is, $\te(\tx)$, and thus, the segment of $\te(\tx)$ between the intersections of $\te(\tx)$ and $\eta_o(\xi_o)$ remains invariable in the three plots. As a result, we find that for the case of $\theta_o=17^\circ$, the blue line and the green line almost coincide in Fig. \ref{asc}, since from the middle plot in Fig. \ref{el}, one can see that the OPR with $r_s=3.01$ coincides with the OPR with $r_s=2.24$ when $\xi<\tx(r_s)$ and only has a tiny difference from the OPR with $r_s=2.24$ when $\xi>\tx(r_s)$. Similarly, the difference between the red and green lines in the case of $\theta_o=17^\circ$ is visible in Fig. \ref{asc} since one can see that the difference in their OPRs is evident from the right plot in Fig. \ref{el}. Moreover, from the right plot in Fig. \ref{asc}, we can see that the difference between the green and blue lines becomes significant on the right, and the three lines are very close in the left part. The reason can be easily found in Fig. \ref{el}, where the opening of the parabola $\eta_o(\xi_o)$ increases when $\theta_o$ goes from $17^\circ$ to $80^\circ$. Furthermore, in the middle plot of Fig. \ref{el}, the difference in the OPRs becomes larger at $\theta_o=80^\circ$, and in the right plot of Fig. \ref{el}, the red and blue lines intersect very closely with the purple line since $r_s=3.92$ is near $r_{p+}=3.56$.

Therefore, qualitatively, we can conclude that when $r_s<r_{p-}$, the shadow of the CO is the same as that of the black hole; when $r_{p-}<r_s<r_{p+}$, the shadow of the CO is larger than that of the black hole, and the shadow of the CO becomes slightly larger as $\theta_o$ increases from $0^\circ$ to $90^\circ$ with parts of the shadow curves overlapping; and when $r_s>r_{p+}$, the shadow of the CO becomes significantly larger, and each point of the CO shadow curve is outside the corresponding end of the black hole shadow curve.

\subsection{Quantitative study of the variation in the CO shadow}

In this subsection, we give a quantitative study of the variation of the shadow concerning the radius of the surface of a CO. Following the work \cite{Zhong:2021mty, Zhang:2022osx}, we use the average radius $\bar{R}$ as the characteristic length of a shadow.

\begin{figure}[h!]

\centering
\includegraphics[width=3.5in]{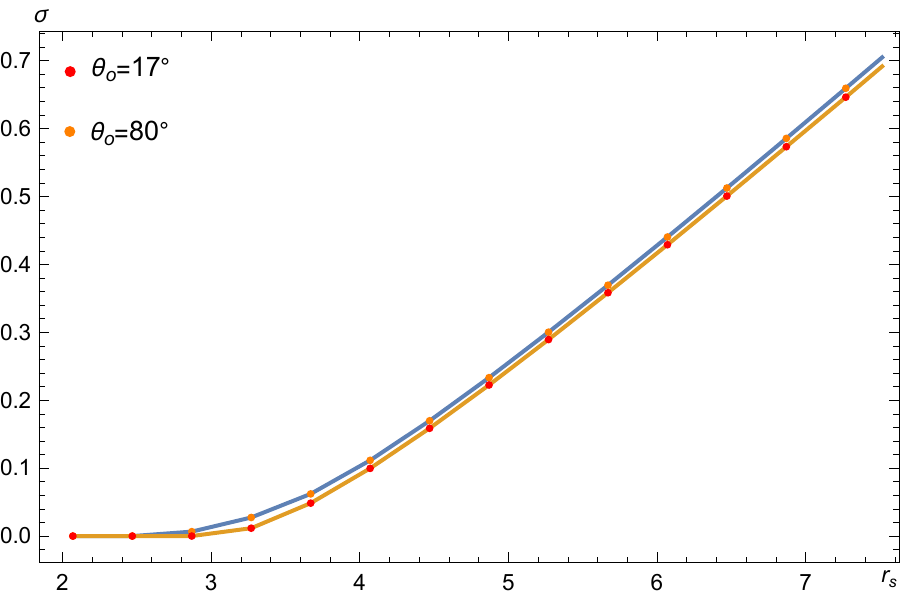}

\caption{The variation in the dimensionless parameter $\sigma=\bar{R}/\bar{R}_0-1$ of the CO shadow concerning the radius of the surface of the CO. In the plot, we set $r_s=2.07+0.4(i-1)$, where $i=1, 2,\dots, 14$ for each point. }\label{sbar}
\end{figure}

In Fig. \ref{dia}, we present a diagram to show the coordinates of points at which the shadow curve intersects two axes. $O$ is the origin of the Cartesian coordinates on the screen. Considering the $\mathcal{Z}_2$ symmetry of the spacetime, the center of the shadow can be defined as $\left(x_c=\frac{\xmi+\xma}{2}, \frac{\ymi+\yma}{2}=0\right)$. Then, let $(x_c, 0)$ be the center, and we can introduce polar coordinates $(R, \psi)$ with $R=\sqrt{(x-x_c)^2+y^2}$. The parameter $\bar{R}$ can be defined as
\bea
\bar{R}=\int_0^{2\pi}\frac{R(\psi)}{2\pi}d\psi\,,
\eea
which denotes the average radius of the shadow curve. It is convenient to introduce a dimensionless parameter
\be
\sigma=\frac{\bar{R}}{\bar{R}_0}-1\,,
\ee
where we use $\bar{R}_0$ to represent the average radius of the shadow curve when $r_h<r_s<r_{p-}$. In Fig. \ref{sbar}, we show the variation in $\sigma$ concerning the radius of the CO surface, where we fix $M=1$ and $J=0.5$ and set $r_s=2.07+0.4(i-1)$ with $i=1, 2, \dots, 14$. We find that the average radius of the shadow curve increases slowly as the radius of the CO surface increases from $r_{p-}$ to $r_{p+}$, because $r_{p+}-r_{p-}=1.09$
is small. When $r_s>r_{p+}$, the average radius of the shadow curve increases quickly as the radius of the CO surface increases, and the change is almost linear. In addition, we can see that the average radius of the shadow curve at $\theta_o=80^\circ$ is always larger than that at $\theta_o=17^\circ$ for a fixed $r_s$ in the range $r_s>r_{p-}$, which agrees well with our analysis in the last subsection.

\section{Summary}\label{sec4}

In this work, we studied the problem of comparing shadows of COs with and without horizons. For simplicity, the CO was considered not to emit or reflect any light compared to other luminous sources in the background of the CO. In addition, we assumed that the CO is a slowly rotating object such that the spacetime outside the surface of the CO can be described by the Painlev\'e-Gullstrand form of the Lense-Thirring metric. In terms of the photon region with $r_{p-}\le\tilde{r}\le r_{p+}$, we investigated three cases, that is, the radius $r_s$ of the CO is smaller than $r_{p-}$, $r_{p-}<r_s<r_{p+}$ and $r_s>r_{p+}$. To obtain the shadow curve for different cases, we introduced OPR and POPR in Sec. \ref{3s1} to construct a clear correspondence between the shadow curve and the impact parameters. Moreover, we recognized a new class of critical impact parameters $\eta_s(\xi_s)$, with which the photons have a turning point at $r_s$. 
After a detailed analysis of the OPRs and POPRs for COs with various $r_s$, we found the POPR governed by the photon region $\te(\tx)$, which is the same as that for black holes when $r_h<r_s<r_{p-}$, one part of the POPR is governed by the photon region $\te(\tx)$, and the other part is controlled by $\eta_s(\xi_s)$ when $r_{p-}<r_s<r_{p+}$; the POPR is completely controlled by the $\eta_s(\xi_s)$ when $r_s>r_{p+}$. 
As a result, compared with the shadow curve of a black hole, we found that the shadow curve of a CO does not change when $r_h<r_s<r_{p-}$, partially changes when $r_{p-}<r_s<r_{p+}$ and completely changes when $r_s>r_{p+}$. We also performed a quantitative study
on the variation of the shadow curve concerning $r_s$ and found that the average radius of the shadow curve increases slowly when $r_s$ goes from $r_{p-}$ to $r_{p+}$ and very quickly when $r_s$ increases after $r_{p+}$.

Our results indicate that a CO with or without a horizon is not distinguished by the shadow curve when it has a whole photon region outside its surface. A CO without a horizon can be distinguished from a black hole when the photon region is partially or entirely hidden in the surface of the CO; that is, in this case, the EHT can be used to determine whether a CO has an event horizon if the resolution reaches high enough. Although in the present work, our discussion is based on an approximate metric, our results should not depend on a specific metric but instead reflect a universal property for a CO. Obviously, further study considering a more realistic model is needed.

\section*{Acknowledgments}
The work is partly supported by NSFC Grant Nos. 12205013 and 12275004. MG is also endorsed by "the Fundamental Research Funds for the Central Universities" with Grant No. 2021NTST13.


\end{document}